\documentclass[twocolumn]{aastex631}

\usepackage{amsmath}

\received{}
\revised{}
\accepted{}

\submitjournal{ApJ}

\shorttitle{DM Measurements}
\shortauthors{Lewandowska et al.}

\begin{document}

\title{Single Pulse Dispersion Measure of the Crab Pulsar}

\correspondingauthor{Natalia Lewandowska}
\email{nlewando@nrao.edu}

\author[0000-0003-0771-6581]{N. Lewandowska}
\affil{Department of Physics and Astronomy, West Virginia University\\ Morgantown, WV 26506, USA}
\affil{Center for Gravitational Waves and Cosmology, West Virginia University, Chestnut Ridge Research Building, Morgantown, WV, 26505, USA}
\affil{Swarthmore College, 500 College Ave, Swarthmore, PA 19081, USA}

\author{P.\,B. Demorest}
\affiliation{National Radio Astronomy Observatory, 1003 Lopezville Road, Socorro, NM 87801, USA}

\author{M.\,A. McLaughlin}
\affil{Department of Physics and Astronomy, West Virginia University\\ Morgantown, WV 26506, USA}
\affil{Center for Gravitational Waves and Cosmology, West Virginia University, Chestnut Ridge Research Building, Morgantown, WV, 26505, USA}

\author[0000-0002-8906-7783]{P.\,Kilian}
\affil{Space Science Insitute, 4765 Walnut St, Suite B, Boulder, CO 80301}

\author{T.\,H. Hankins}
\affiliation{Physics Department New Mexico Tech, Socorro, NM 87801, USA}
\affiliation{Adjunct Astronomer at National Radio Astronomy Observatory}

\begin{abstract}
We investigate the use of bright single pulses from the Crab pulsar to determine separately the dispersion measure (DM) for the Main Pulse and Interpulse  components.
We develop two  approaches  using cross correlation functions (CCFs). The first method computes the CCF of the total intensity of each of 64 frequency channels with  a reference channel and converts the time lag of maximum correlation into a DM. The second method separately computes the CCF between every pair of channels for each individual bright pulse and extracts an average DM from the distribution of all channel-pair DMs. Both methods allow the determination of the DM with a relative uncertainty of better than 10$^{-5}$ and provide robust estimates for the uncertainty of the best-fit value. 
We find differences in DM between the Main Pulse, the Low
Frequency Interpulse, and the High Frequency Interpulse using both
methods in a frequency range from 4 to 6 GHz. Earlier observations of the High Frequency Interpulse carried out by \cite{hankins_2016} resulted in  DM$_{\rm HFIP}$-DM$_{\rm MP}$ of 0.010 $\pm$ 0.016 pc cm$^{-3}$. Our results indicate a DM$_{\rm HFIP}$-DM$_{\rm MP}$ of 0.0127 $\pm$ 0.0011 pc cm$^{-3}$ (with DM$_{\rm comp}$ being the DM value of the respective emission component), confirming earlier results with an independent method. During our studies we also find a relation between the brightness of single pulses in the High Frequency Interpulse and their DM. 
We also discuss the application of the developed methods on the identification of substructures in the case of Fast Radio Bursts.
\end{abstract}


\section{Introduction} \label{sec:intro}
Since the discovery of the first pulsar in 1967 \citep{hewish_1968}, pulsars and other classes of neutron stars have been studied in great detail at radio wavelengths and beyond \citep{harding_2013}. 
Continuous studies of their properties have not only increased our understanding of  pulsar emission mechanisms, they are also suitable sources for studies of the interstellar medium (ISM). 
{The radio pulses, propagating through the ISM, are subject to  scattering, scintillation \citep{rickett_1990}, Faraday rotation \citep{melrose_1979} and  dispersion} \citep{hewish_1968, tanenbaum_1968, swanson_2003, chen_2012, fitzpatrick_2014}.  
The free electrons in the ISM delay lower frequencies of a broadband pulse compared to the higher frequency part of its spectrum. For a cold plasma the time delay between the reception of a signal at \textit{t(f)} at a frequency \textit{f} and the reception \textit{t(f=$\infty$)} at its inversely proportional to the square of observing frequency, \textit{f}:

\begin{equation}\label{delay}
 t(f) - t(f={\infty}) = \mathcal{D} \cdot \frac{\mathrm{DM}}{f^{2}}
\end{equation}

The quantity $\mathcal{D}$ consists only of natural constants and is known as the \textit{dispersion constant}. The value we have used here is $\mathcal{D}$ = (4.148808 $\pm$ 0.000003)$\cdot$ 10$^{3}$\,MHz$^{2}$ pc$^{-1}$ cm$^{3}$ s \citep{handbook_2012}. The other quantity labelled as DM is the integrated column density of free electrons along the line of sight and is known as \textit{dispersion measure}. It is usually expressed in units of pc cm$^{-3}$  \citep{stix_1962,handbook_2012}.
The free electron density is neither homogeneous, nor constant in time (this can be only assumed as a crude approximation, \citealt{handbook_2012}), but differs in concentration  \citep{cordes_2002}\footnote{https://www.nrl.navy.mil/rsd/RORF/ne2001/}. DM measurements of pulsars are used for measuring the column density of free electrons \citep{yao_2017}.

Numerous DM studies based on pulsar data have been carried out in the past to examine the validity of the cold plasma dispersion law,  Equation~\ref{delay} \citep{tanenbaum_1968,phillips_1992}. Multifrequency observations carried out by \cite{phillips_1992} result in the conclusion that integrated profile arrival times follow the cold plasma dispersion law at meter and decameter wavelengths. Departures are observed for two pulsars in their sample at frequencies from 2 to 5 GHz.
In addition, temporal changes of the DM have been extensively observed and studied \citep{goldstein_1969,rankin_1971,hankins_1987,hankins_1991,phillips_1991,backer_1993}, addressing ordinary as well as recycled pulsars \citep{rawley_1988,cognard_1997}. Frequency-dependent DM variations were also discovered in the case of recycled pulsars \citep{ramanchandran_2006}.

Temporal changes of DM values can be caused by discrete structures in the ISM, changes of the relative pulsar velocity and the Earth as well as changes of the solar wind \citep{lam_2016}.
A study \citep{jones_2017}, of DM variations in 37 recycled pulsars shows periodic annual, monotonically increasing or decreasing trends, or both. Stochastic trends are reported too. The DM fluctuations have timescales  from two weeks up to more than one year, suggesting the existence of discrete structures in the ISM.

Frequency-dependent DM caused by fluctuations in the electron column density of
the ISM  have also been observed for the first detected recycled pulsar PSR B1937+21 \citep{ramanchandran_2006}. Such variations are assumed to be caused by a turbulent medium and the effect of multipath propagation. They deduce that frequency-dependent DM variations can have a significant impact on the long term timing accuracy of a pulsar.
A theoretical approach for  frequency dependent DM has benn developed by \cite{cordes_2016}, explaining it as a result of multipath scattering caused by inhomogeneities of the ISM containing different electron densities. 

A standard approach to determine a pulsar DM  is by measuring the time delay (Equation~\ref{delay}) between pairs of average profiles recorded at wide bandwidths \citep{hankins_1987,hankins_1991}. As described in \cite{hankins_1987} depending upon the profile of a pulsar there are several methods to achieve the alignment of average profiles. 
Profile evolution with frequency \citep{komesaroff_1970} can lead to frequency dependent measurements of DM. 
An independent approach to determine the DM was developed using  microstructure \citep{craft_1968} discovered in some pulsars.  The correlation between pairs of single pulse microstructure  at different frequencies was examined \citep{hankins_1991}. \color{black} An interesting difference between both approaches is a lower DM value resulting from the cross correlation of microstructure than from profile alignment. However, the results from different studies using microstructure determined DM are not all consistent (see \citealt{hankins_1991}, and references therein), some resulting in low frequency delays when used for the alignment of profiles and therefore suggesting the existence of \emph{superdispersion} \citep{shitov_1988}. These studies suggest the use of single pulses for the determination of the DM of a pulsar as an alternative approach to the alignment of profiles due to the complications caused by profile evolution. However, they do not state which method produces the more accurate values \citep{hankins_1991}.
Further DM studies carried out with profiles as well as with single pulses are reported by \cite{ahuja_2005} who monitored several pulsars with dual frequency observations for over a year. Also in this case, a lower DM resulting from cross-correlating single pulses was observed, albeit not in all pulsars in their sample. Stating that a pulsar's DM depends on the method of the analysis, they also emphasize that both methods they apply measure different quantities. While in both cases cross-correlation functions (CCFs) are used, the CCF in the case of profiles describes the cross-correlation of the sum of all pulses from different radio (sub)bands. An average profile is built by folding a time series at the respective pulsar period. Hence a profile component is much broader than a single pulse which is also reflected in a broader CCF resulting from profiles than from single pulses. For single pulses the CCF is the sum of all cross-correlations between the single pulses from two radio bands. Since the CCF is a bi-linear operator the two approaches are not necessarily the same. \color{black} \citep{ahuja_2005}.

\begin{figure*}
\includegraphics[width=\textwidth]{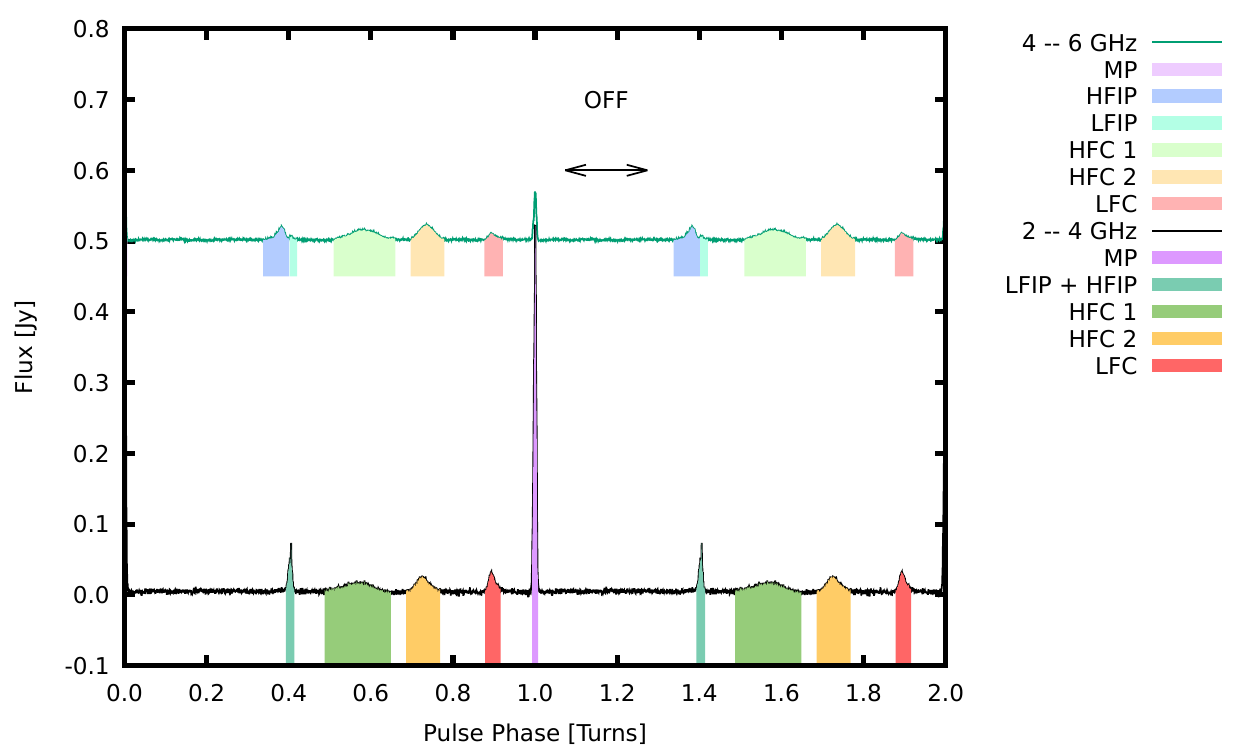}
\caption{
Lower curve: 2 - 4\,GHz profiles, upper curve: 4 - 6\,GHz profiles. The dedispersion was carried out with DM$_{\rm JB}$ = 56.778 pc cm$^{-3}$. The Main Pulse (MP) is located at phase 0, 1 and 2. The Low Frequency Interpulse (LFIP) is at 0.4 and 1.4 in phase. The High Frequency Components (HFC) 1 and  2 are located in the phase range from about 0.5  to 0.75 (1.5  to 1.75 respectively) and the Low Frequency Component (LFC) is seen at a phase of about 0.9 and 1.9. The OFF pulse statistics are calcuated in a phase range from 1.07 to 1.27. The colored areas indicate rotational phase ranges used for our Gaussian fits (Table~\ref{prof_fits}).\label{pro}}
\end{figure*}

Here we choose the approach of cross-correlating single pulses for a determination of the true DM for multiple frequencies of our  data set. Single radio pulses are sharper in time than average profiles. This characteristic results in a sharp peak of the corresponding CCF.
Since its detection by sporadic single pulses \citep{staelin}, which were later labelled as giant pulses (GPs), the Crab pulsar has been the subject of numerous radio studies. High time resolution studies revealed the existence of seven average radio emission components \citep{moffett_hankins_1996, hankins_2007, hankins_2015}. At low frequencies the \textit{Main Pulse} (MP) emission component is by far the strongest average profile component   (Figure~\ref{pro}), but it vanishes above frequencies of 5\,GHz \citep{hankins_2015}. However, MP single pulses have been observed up to 43\,GHz \citep{hankins_2016}. The Crab pulsar possesses two interpulse components: the \textit{Low Frequency Interpulse} (LFIP) which is dominant at frequencies below 5\,GHz and a second interpulse component known as \textit{High Frequency Interpulse} (HFIP) which occurs about 7 degrees earlier in rotational phase and is dominant at frequencies above 5\,GHz \citep{moffett_hankins_1996}.  So far, such complex behavior regarding the average emission components has not been observed in any other pulsar. Single pulse studies revealed a higher DM value for the HFIP component than the MP single pulses,  
indicating possible  propagation effects in the pulsar magnetosphere \citep{hankins_2007}. Apparent differences in DM among pulsar profile emission components suggests that they originate at different heights in the pulsar magnetosphere. However, it is difficult to explain which plasma dispersion law they follow since the cold un-magnetized plasma dispersion law might not necessarily be valid in a pulsar magnetosphere \citep{eilek_2016}. 

The data used for the present study were recorded with the Karl G. Jansky Very Large Array (VLA). In phased-array mode the synthesized beam is very narrow, 
which makes it possible to resolve and thereby reject the background resulting from the extended Crab Nebula. 
Consequently, weaker single pulses from the Crab pulsar can be detected than with a single-dish radio telescope. 
The recent upgrade of the VLA has provided new capabilities for continuous sampling and wider bandwidths.
Here we examine the DM of the Crab pulsar emission components. We determine the DM for three out of the detectable six emission components in our data (the low-frequency precursor is not seen above about 1\,GHz), using their single pulse emission with two different techniques and discuss  the results in view of current pulsar and Fast Radio Bursts (FRB) studies. 
Section~\ref{radio_obs} describes our observations. In Section~\ref{dm_var} the DM  measurements are described. The error analysis is described in Section~\ref{err_ana}.  
The results of our analysis are discussed in Section~\ref{comp_dm}. An application of the results to our data is described in Section~\ref{appl}. 
Our finding of two HFIP single pulse populations is described in Section~\ref{single_pulses_HFIP}.
The results of our methods and their significance are discussed in Section~\ref{disc}. A summary of our results together with the corresponding conclusions is given in Section~\ref{conc}. 

\section{Observations \& Data Reduction} \label{radio_obs}
Our observations  were recorded with the VLA in D-configuration, using 27 antennas in phased-array mode at S--band (2 - 4\,GHz) and C--band (4 - 6\,GHz). The data were recorded in baseband format and reduced offline. For coherent dedispersion \citep{hankins_rickett_1975} the digital library DSPSR \citep{van_straten_2011} was used with an initial DM$_{\rm JB}$ of 56.778 pc cm$^{-3}$ provided by the Jodrell Bank Observatory \citep{lyne_1993}. We determined the residual $\Delta$DM = DM$_{\rm JB}$ -- DM  by our CCF methods. We do so for each of the different emission components of the Crab pulsar separately and refer to the resulting value as $\Delta$DM$_{\rm comp}$. For plotting, e.g., Figure~\ref{data_applic}, we reprocessed the baseband data with that $\Delta$DM$_{\rm comp}$.
To remove some of the uncertainty in the determination of the true DM of the different emission components of the Crab pulsar, we referenced all resulting $\Delta$DM$_{\rm comp}$ to the DM of the MP in that frequency band (see Table~\ref{delta_dm_components}) using $\delta$DM$_{\rm comp}$ = $\Delta$DM$_{\rm comp}$- $\Delta$DM$_{\rm MP}$.

The extraction of single pulses was carried out using routines from the PSRCHIVE software package \citep{van_straten_2012}. To examine only bright single pulses, an intensity threshold of 5$\sigma$  (which corresponds to five times the rms offpulse noise) 
was introduced during the extraction. 
The number of occurrences of single pulse components exceeding the 5$\sigma$ threshold are shown in Table~\ref{radio_sum}. All further calculations were carried out using the PSRCHIVE Python Interface\footnote{http://psrchive.sourceforge.net/manuals/python/} \citep{van_straten_2012}.

\begin{table*}
\tablenum{1}
\begin{center}
\begin{tabular}{l r r}
Epoch  & 2016 Jan 06 & 2016 Jan 07\\
Center Frequency (MHz) & 5000 & 3000\\
Bandwidth (MHz)  & 2048 & 2048\\
Number of Channels  & 64 & 64\\
Duration (Number of Crab Pulsar Periods) & 84000 & 84364\\
Number of Single Pulses (MP) & 7970 & 27157\\
Number of Single Pulses (LFIP + HFIP) & \dots  & 23582\\
Number of Single Pulses (HFIP) & 12118  & \dots\\
Number of Single Pulses (LFIP)  & 1196 & \dots\\
\end{tabular}
\caption{Summary of Observations. \label{radio_sum}}
\end{center}
\end{table*}

\section{Dispersion Measure Determination} \label{dm_var}
A pulsar detection often employs trial dedispersion (see \citealt{handbook_2012}, Chapter 6.1) which can leave some residual uncorrected dispersion. 
Here  we therefore examine a method how precisely the remaining DM after initial dedispersion, referred to as $\Delta$DM, of single pulses can be determined.

$\Delta$DM is determined here by two different techniques. Dispersion of pulsar signals delays their time of arrival at lower frequencies to later times.
The CCF enables the determination of the time shift \citep{knapp_1976,azaria_1984} without requiring knowledge of details of the underlying plasma models, such as the assumption of a cold plasma or a negligibly low plasma frequency (see discussion in Chapter 4.2 in \citealt{handbook_2012}). 
A CCF of the two channels named $s_1$ and $s_2$ which have a time delay of $\tau$ between each other, is calculated the following way:

\begin{equation}\label{ccf}
 CCF(\tau) = \int s_1(t) \cdot s_2(t-\tau)\,\mathrm{d}t
\end{equation}

We determine CCFs between dedispersed, total intensity 
signals $s_1$ (at frequency $f_1$) and $s_2$ (at $f_2$) coming from each
of the 64 recorded frequency channels.
Our methods differ in the way  the permutations between the channels are determined; (1)
 by holding one reference channel fixed and determining the CCFs between the reference channel
 and all other channels (Method 1, Section~\ref{least_sqa}). (2) by  determining the CCFs between all pairs of channels (Method 2, Section~\ref{freq_time_lag}). For Method 2, Equation~\ref{ccf} is integrated over a short time span around a bright single pulse, the $\Delta$DM is determined via Equation~\ref{dis_constant}, and the result is included in a histogram as shown in Figure~\ref{histo1}. Both approaches are described in detail in the following
 section.

\subsection{Method 1: Fixed Reference Frequency} \label{least_sqa}
For this method we define the channel with the highest frequency in the data set as the reference channel, or \emph{$chan_1$}. A 5$\sigma$ threshold is applied to the signal $s_1$ at the reference frequency $f_1$.  
Depending upon which emission component is being chosen, all other pulse phase ranges in the file are excluded, or masked.  
We then pick one of the other 63 frequency  channels and refer to it as \textit{$chan_{j}$} at frequency $f_{j}$ with  \textit{j} going from 2 to 64 in this iteration.
This procedure is carried out for each of the 63 other frequency channels in the data set in turn which are uniformly referred to as \emph{$chan_j$}. 
We calculate the CCF in steps of the recorded time resolution $\Delta$\textit{t} = 1$\mu$s for -100\,$\Delta t \leq \tau \leq$ 100\,$\Delta t$.  
Outside this range the CCF falls to negligible values for all frequency pairs. 
To be precise, we calculate a discrete, zero mean, normalized cross correlation defined by 

\begin{align}\label{ccf2}
&\mathrm{CCF}_{ij}(\tau) =\\ &\sum_{n=1}^{N_{\mathrm{s}}}
\frac{(s_i(n\Delta{t}) - \left<s_i\right>) \cdot (s_j(n\Delta{t} - \tau) - \left<s_j\right>)\cdot w(\varphi)}{\sigma_i\,\sigma_j\,N_{\mathrm{s}}} \quad.\nonumber
\end{align}

The quantity $N_{S}$ is the number of time samples, the quantities with $\langle \rangle$ representing the mean which are necessary to subtract offsets and the quantities shown by a $\sigma$ are the standard deviations used for normalization of the signal. We define
w($\varphi$) as a window function  for the emission component of interest and zero for all other pulse phases $\varphi$. The center and width of the windows used for the different emission components are given in Table~\ref{prof_fits}. Inside the CCF computation $\varphi$ is given by the fractional part of $i\Delta{t}/P$. 
After calculating the normalized CCF$_{1j}(\tau)$ for all \emph{chan\_j} and all $\tau$ values, the peak of the CCF (for a specific frequency value $f_2$) is searched for along the time axis.
The determination of $\Delta$DM$_{\rm comp}$ is then carried out by fitting Equation~\ref{delay} to the time delays.
The results of this technique are shown in Figures~\ref{data_applic} and ~\ref{data_applic_II} and summarized in Table~\ref{method1} for the MP, HFIP and LFIP emission  components in both frequency bands. We split the analysis into two stages: the determination of $\Delta$DM$_{\rm comp}$ resulting from the data dedispersed with DM$_{\rm JB}$ (Stage 1) and the calculation of $\Delta$DM$_{\rm comp}$ after reprocessing the data with the DM corrections resulting from Stage 1 (Stage 2). We note that the results shown in Table~\ref{method1} are obtained during Stage 1 of the analysis. The results shown in 
Figures~\ref{data_applic} and ~\ref{data_applic_II} are obtained after correcting the data with the respective $\Delta$DM$_{\rm comp}$ as described in Section~\ref{appl}, hence in Stage 2 of the analysis. We also point out that the blank areas in Figures~\ref{data_applic} and ~\ref{data_applic_II} are frequency channels that have been affected by Radio Frequency Interference (RFI) and were excised.

\begin{deluxetable*}{ccccc}
\tablecaption{Summary of residual DM values (labelled as $\Delta$DM$_{\rm comp}$) obtained with both methods.\label{method1}}
\tablenum{2}
\tablehead{
\colhead{Emission Component} & \colhead{$\Delta$DM$_{\rm comp}$} & \colhead{FWHM($\Delta$DM)$_{\rm comp}$} & \colhead{Method}  & \colhead{Frequency Band}\\
\colhead{}                   & \colhead{[pc cm$^{-3}$]}          & \colhead{[pc cm$^{-3}$]}                & \colhead{}        & \colhead{[GHz]}
}
\startdata
MP & 0.0182 & 0.0003 &  1 & 2 -- 4\\
LFIP + HFIP & 0.0193 &  0.0003 &  1 & 2 -- 4\\
HFIP & --0.79 & 0.23 & 1 & 3.5 -- 4\\
LFIP & 0.0196 & 0.0006 & 1 & 3.5 -- 4\\
MP & 0.0184 &  \shortstack{--0.0032\\0.0032} &  2 & 2 -- 4\\
LFIP + HFIP & 0.0182 & \shortstack{--0.0034\\0.0057}  & 2 & 2 -- 4\\
\hline
MP & 0.0236 & 0.0042 &  1 & 4 -- 6\\
HFIP & 0.0089 & 0.0011 & 1 & 4 -- 6\\
LFIP & 0.0207 & 0.0013 &  1 & 4 -- 6 \\
MP & 0.02 & \shortstack{--0.01\\0.02} &  2 & 4 -- 6\\
HFIP & 0.02 & \shortstack{--0.06\\0.03} & 2 & 4 -- 6\\
LFIP & 0.0196 & \shortstack{--0.0034\\0.0034} & 2 & 4 -- 6\\
\enddata
\end{deluxetable*}

\subsection{Method 2: Dispersive Time Lag of Single Bright Pulses} \label{freq_time_lag}
The approach described here varies from the one described in Section~\ref{least_sqa} by determining the CCF between all pairs of channels (using CCF$_{ij}$($\tau$)) and the same time lags $\tau$ as in Method 1. In addition we do not integrate over the full observation
time \emph{T}, but only over a short time surrounding each single pulse. 
We determine the time lag $\tau_{\mbox{\rm peak}}$ by fitting the peak of the CCF with a parabola and converting the resulting time value to a DM using Equation~\ref{dis_constant} (from \citealt{handbook_2012}):

\begin{equation}
 \Delta DM = \frac{\tau_\mathrm{peak}}{4.148808 \cdot 10^6 \,\mathrm{s} \cdot (f_{1}^{-2} - f_{2}^{-2})\,\mathrm{MHz}^{2}} \, \frac{\mathrm{pc}}{\mathrm{cm^{3}}} \label{dis_constant}
\end{equation}

\begin{figure}
\includegraphics[width=\columnwidth]{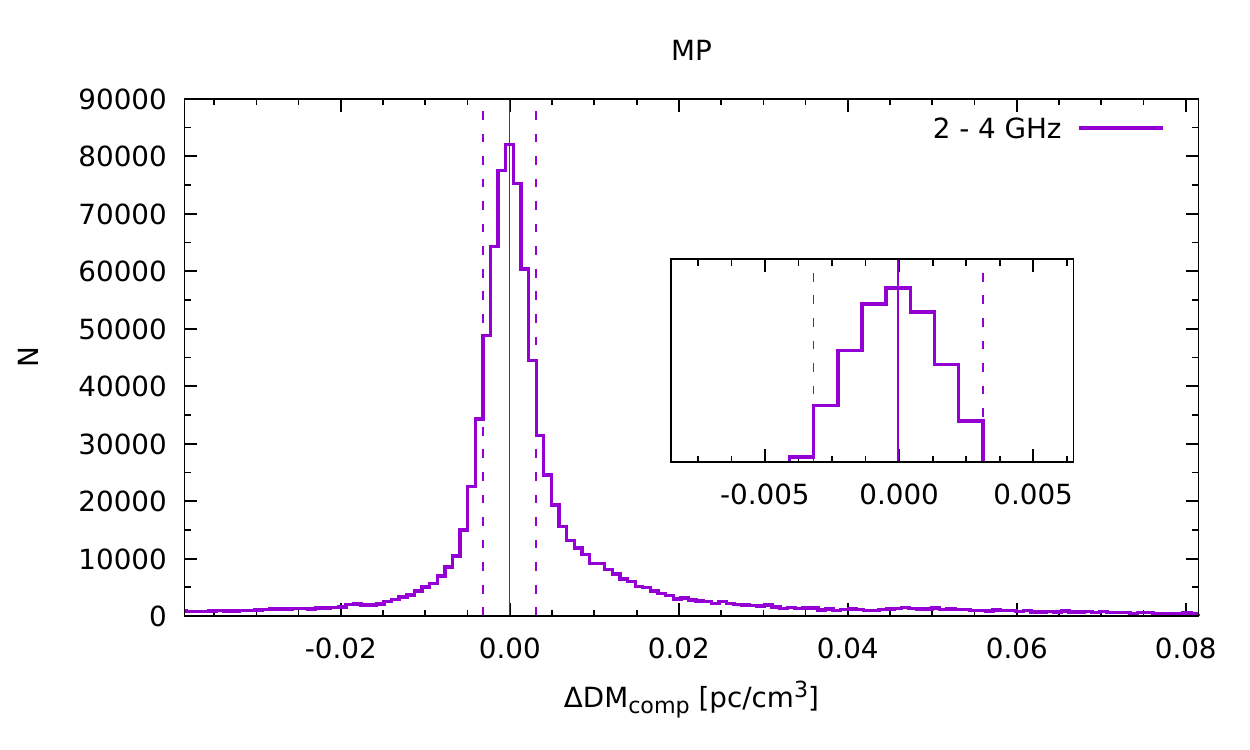}
\includegraphics[width=\columnwidth]{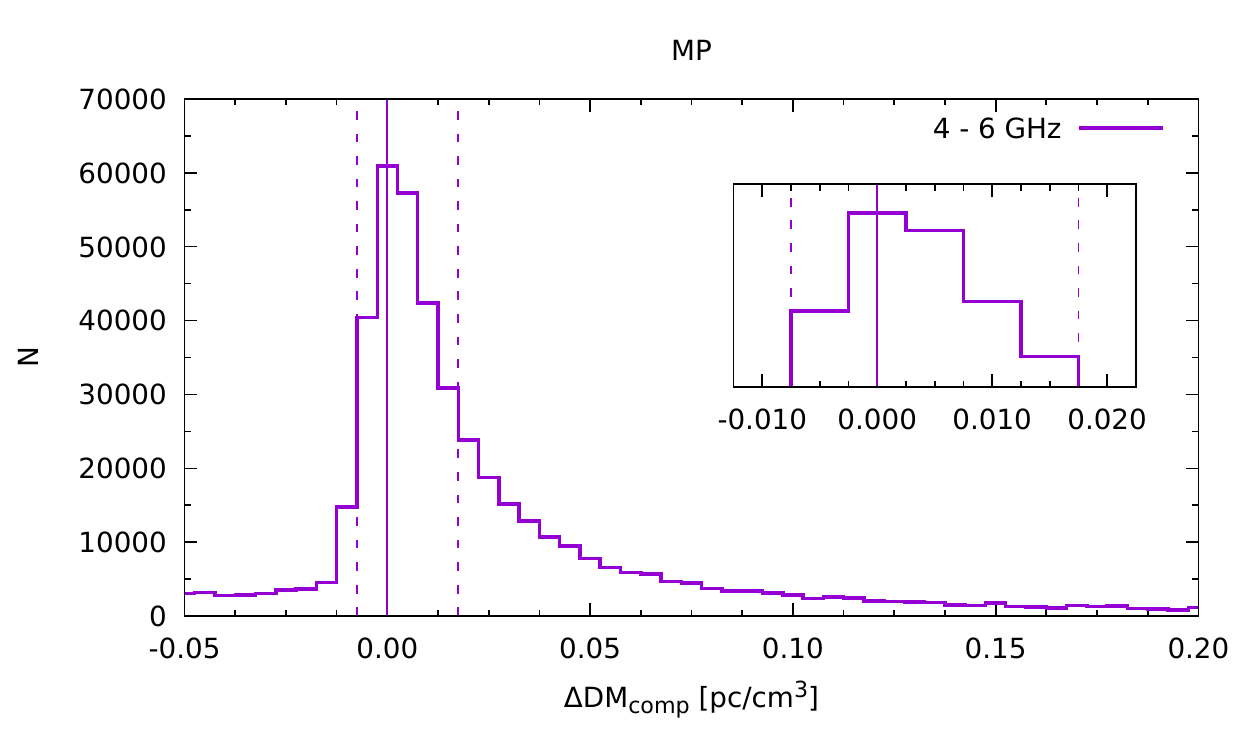}
\caption{The distribution of $\Delta$DM values for the MP emission component in the 2 -- 4\,GHz (top) and 4 -- 6\,GHz (bottom) band. The solid line corresponds to the absolute peak of the histogram and the dashed lines with the FWHM. The quantity $N$ stands for the number of events per bin. \label{histo1}} 
\end{figure}

We produce histograms of the individual DM values, as shown in Figure~\ref{histo1} for the MP emission components in both bands. In both cases the peak of the distribution is not centered around a $\Delta$DM value of zero, indicating the existence of remaining dispersion in the data. 
The resulting $\Delta$DM$_{\rm comp}$ values with the embedded full width at half maximum (FWHM) in Figure~\ref{histo1} are included in Table~\ref{method1}.

\section{CCF Error Analysis}\label{err_ana}\label{err_ana_I}
The uncertainty of Method 1 is assessed by applying a Least Squares fit \citep{legendre1805nouvelles} to the curve expressed by Equation~\ref{dis_constant}. The results from an unweighted Least Squares fit are shown in Table~\ref{method1}. 
The uncertainty estimate for Method\,2 was carried out based on the distribution of $\Delta$DM$_{\rm comp}$ values  shown by the corresponding histograms in Figure~\ref{histo1}. For the estimate of the corresponding uncertainties we use the FWHM which is independent of the distribution. All results are shown in Table~\ref{method1}.

\section{Test for Component-dependent Dispersion Measure} \label{comp_dm}
To test for emission-component dependent DM we carried out the calculations explained in Section~\ref{least_sqa} and \ref{freq_time_lag} for the single pulses occurring at the rotational phase ranges of the MP, HFIP and LFIP \color{black} emission component shown in Figure~\ref{pro}. To determine which single pulses belong to what average emission component in terms of rotational phase range, the latter were fitted with a Gaussian profile convolved with an exponential ISM scattering tail.
The Gaussian is described by its  amplitude $a$, peak phase $\mu$ and width $w$ (details are given in Table~\ref{prof_fits}).  
The exponential tail of a delta impulse is completely described by a scattering time $t_\mathrm{sc}$. We calculated the scattering tails for all emission components in both frequency bands and all values are at or below 10 \,$\mu$s except for the LFC and the HFC components which we do not analyze further here. For the full analytic expression of this convolution we refer to Equation 4 in \cite{mckinnon_2014}. 
The colored areas in Figure~\ref{pro} indicate the fitted rotational phase ranges.
The results of the corresponding $\Delta$DM$_{\rm comp}$ calculations are given in Table~\ref{method1}.

\begin{figure}
\includegraphics[width=\columnwidth]{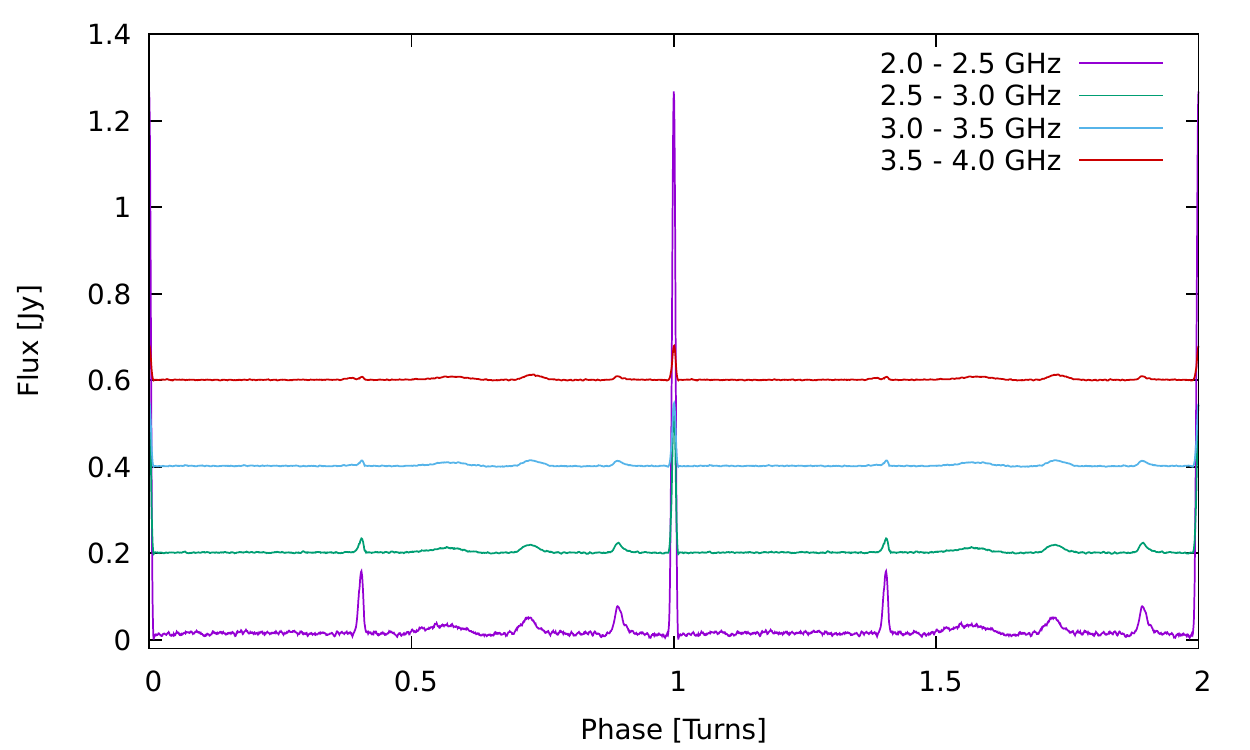}
\caption{Average profiles of the 2 -- 4\,GHz data set split into quarters of the band for the search of the occurrence frequency of the HFIP in this band. This is a smoothed version of the original profile. All curves apart from the one indicating the 2.0 to 2.5 GHz profile are shifted by multiples of 0.2 Jy. \label{profile_S_band_smoothed}}
\end{figure}

\begin{figure}
\includegraphics[width=\columnwidth] {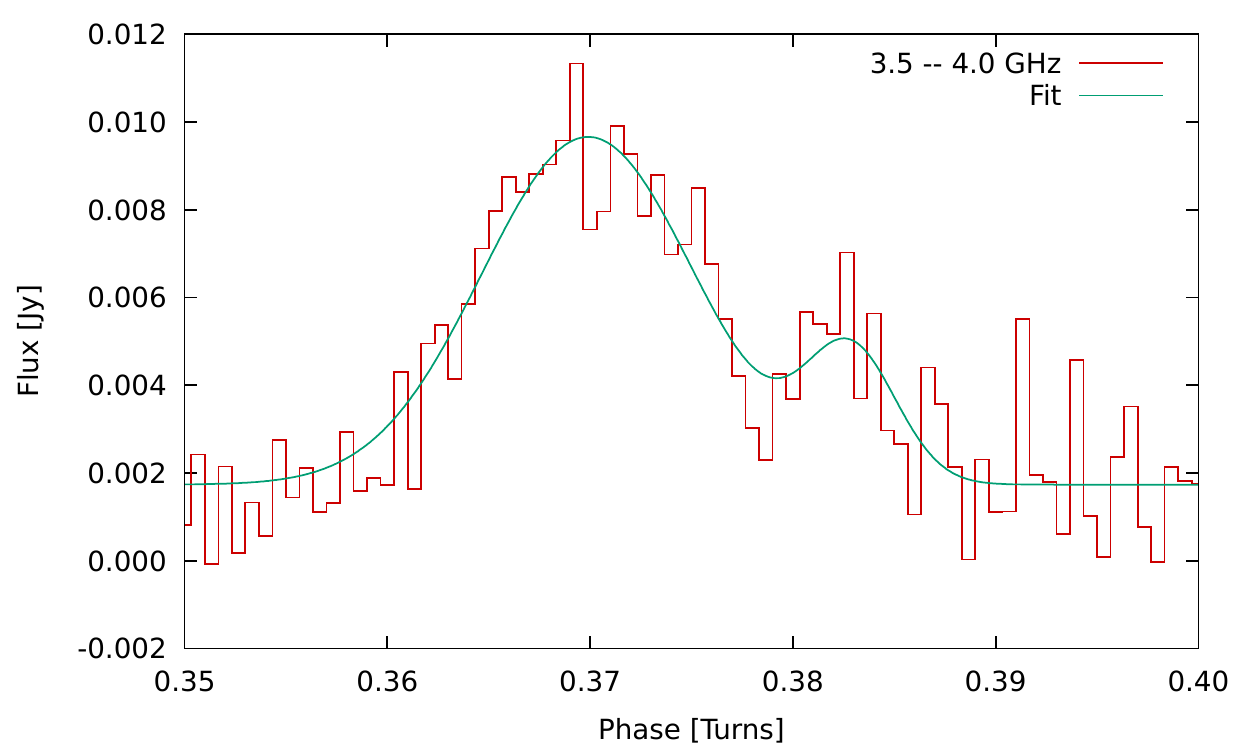}
\caption{Both interpulses in the 3.5 to 4.0 GHz band fitted with two Gaussians and one constant background fit. Left peak: HFIP; right peak: LFIP. 
\label{ip_fit}}
\end{figure}

With our calculations in the 4 -- 6\,GHz frequency band we confirm previous results reported by \cite{hankins_2007} and \cite{hankins_2016}: a higher value of DM for the HFIP compared to the MP and LFIP in that band. 
To examine the frequency at which the HFIP starts occurring as an average emission component in more detail, we split the 2 -- 4\,GHz frequency band into quarters ranging from 2 to 2.5\,GHz, 2.5 to 3\,GHz, 3 to 3.5\,GHz and 3.5 to 4\,GHz (Figure~\ref{profile_S_band_smoothed}). In all four subbands the profile peaks were normalized to the absolute peak of the MP in the respective band. 
We fitted both interpulses as shown in Figure~\ref{ip_fit}. According to our calculations they are located 7.25 degrees away from each other which is in accordance with earlier results \citep{moffett_hankins_1996}.
We determine that the HFIP emission component starts to occur as an average profile component at a frequency of about 3.5\,GHz (Figure~\ref{profile_S_band_smoothed}). 
The $\Delta$DM$_{\rm comp}$ values for both interpulses at frequencies from 3.5 to 4\,GHz are also shown in Table~\ref{method1}. The respective uncertainty of the HFIP $\Delta$DM$_{\rm comp}$ value indicates either a lack of enough data for such a calculation, or poor signal strength. The corresponding $\Delta$DM$_{\rm LFIP}$ shows more resemblance with $\Delta$DM$_{\rm LFIP + HFIP}$ component determined for the entire band from 2 to 4\,GHz, indicating the dominance of the LFIP signal in this part of the band.

\section{Application on Radio Data}\label{appl}
\begin{figure}
\includegraphics[width=\columnwidth]{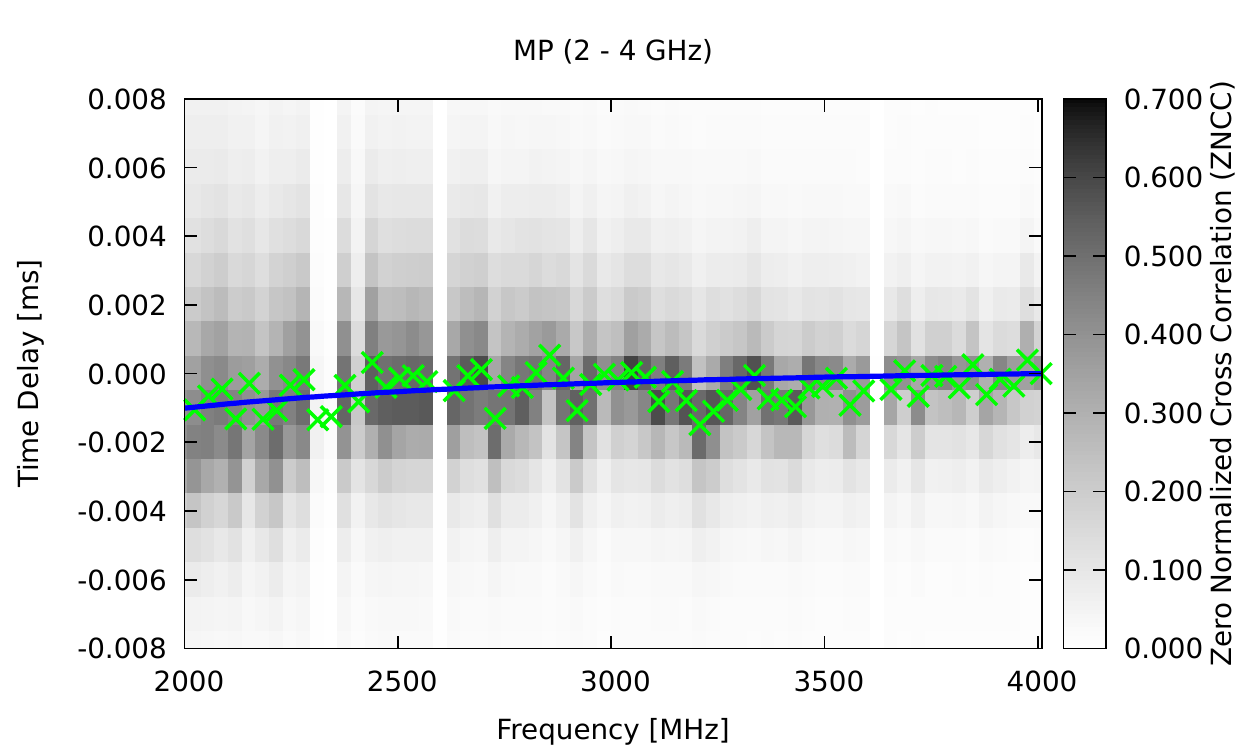}
\includegraphics[width=\columnwidth]{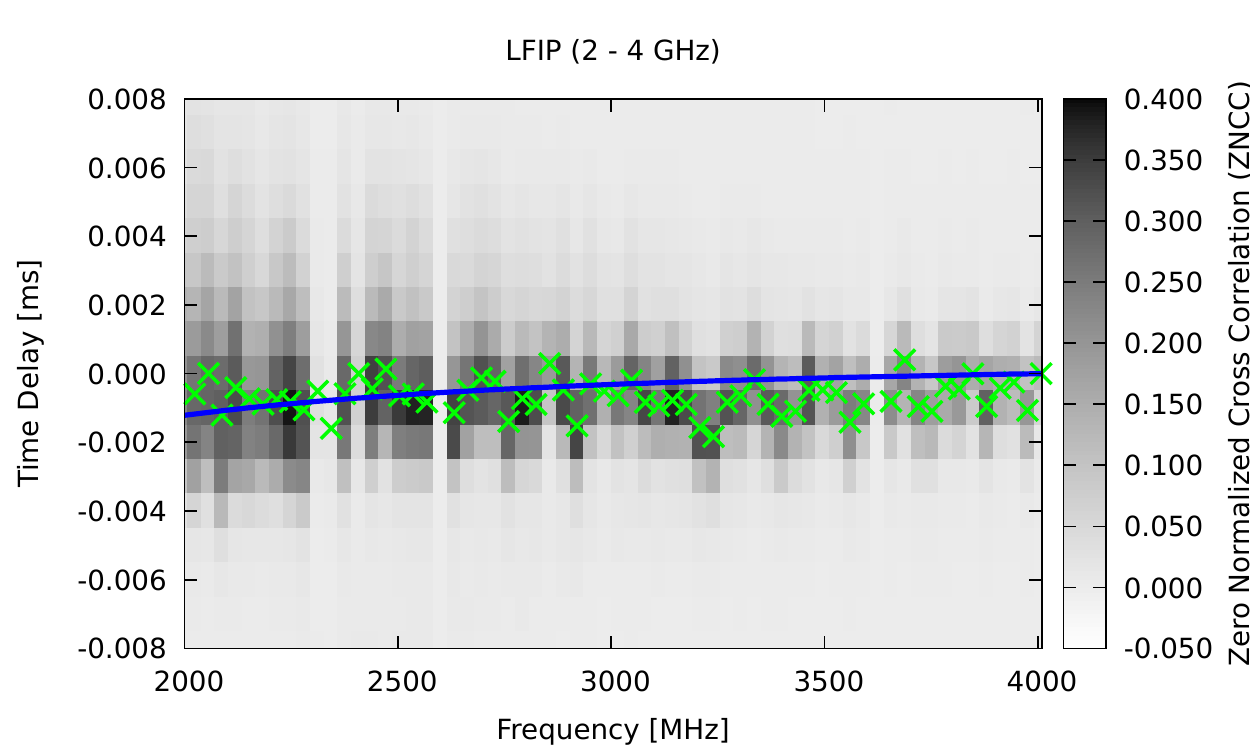}
\caption{ Dedispersed Crab pulsar radio data. The green crosses show the peak of the CCF at a frequency $f$ with reference to $f_{1}$ = 2\,GHz. The blue line shows the best dispersion curve. Top plot: data from single pulses at the phase range of the MP dedispersed using DM$_{\rm JB}$ - $\Delta$DM$_{\rm MP}$; bottom plot: single pulses from the phase range of the LFIP dedispersed using DM$_{\rm JB}$ - $\Delta$DM$_{\rm LFIP}$
All data sets are at 2 - 4 GHz with $\Delta$DM values listed in Table 2.\label{data_applic}}
\end{figure}

\begin{figure}
\includegraphics[width=\columnwidth]{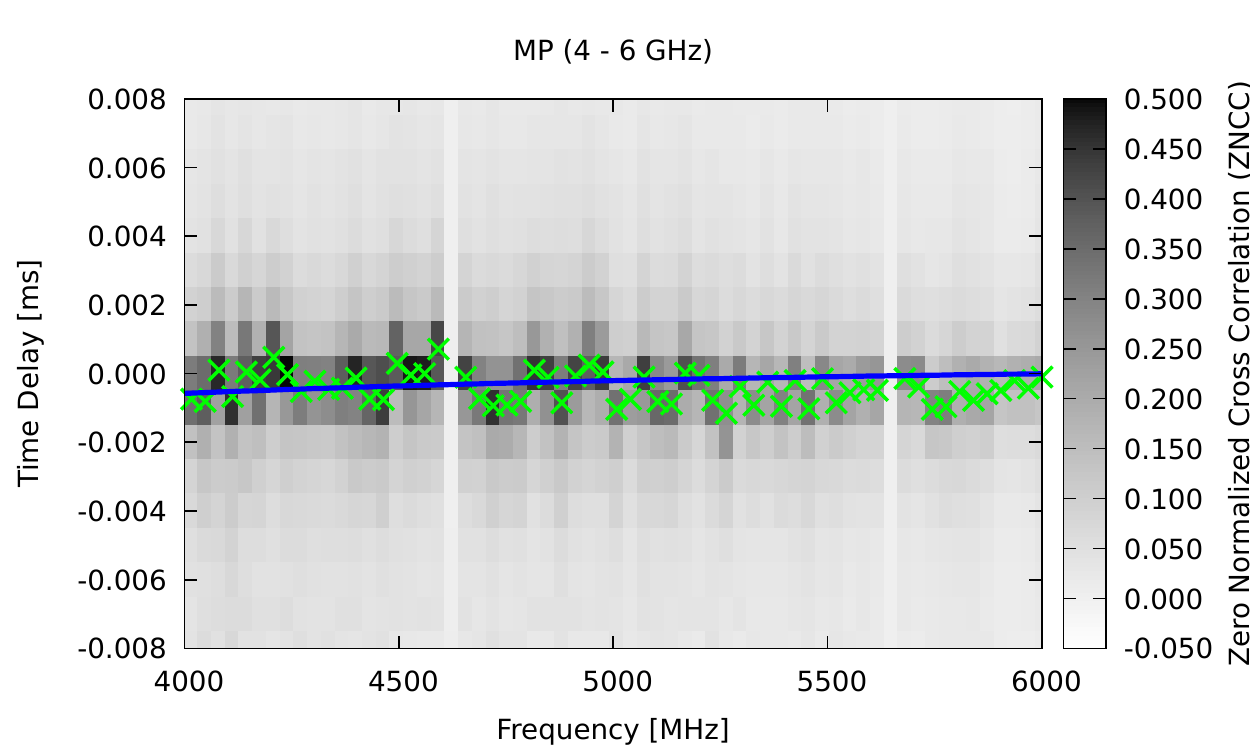}
\includegraphics[width=\columnwidth]{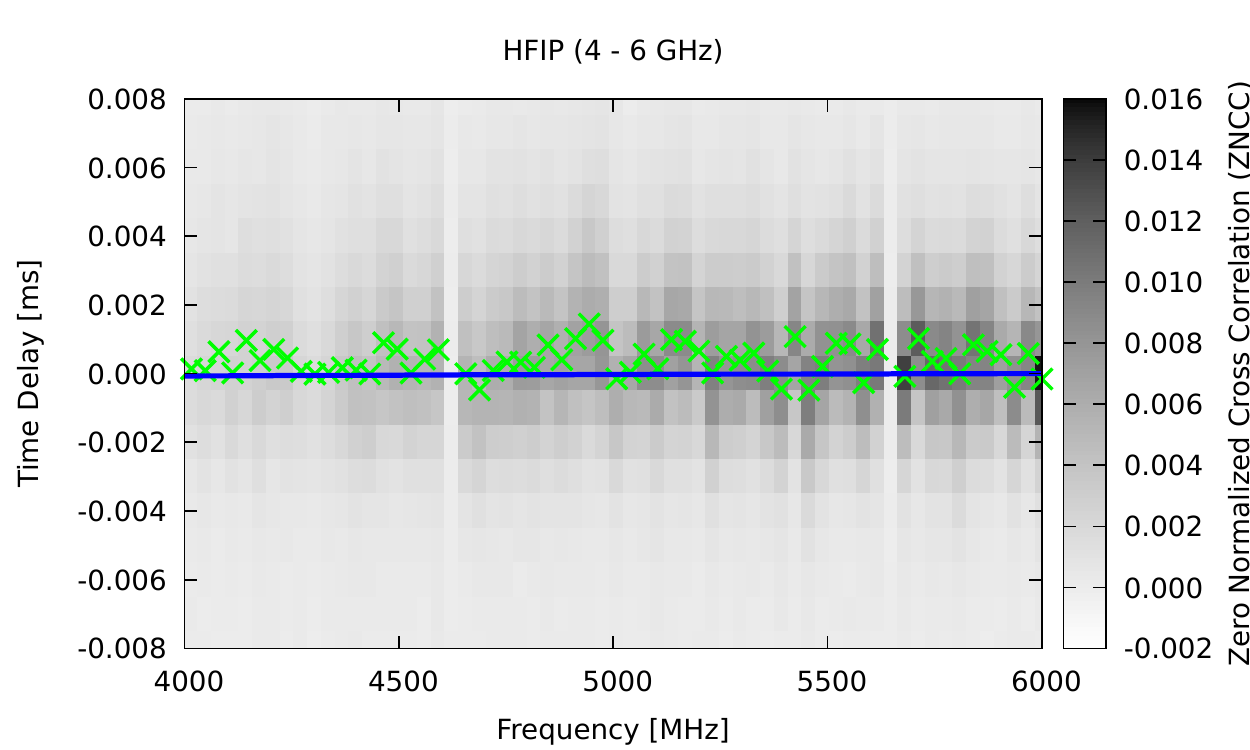}
\includegraphics[width=\columnwidth]{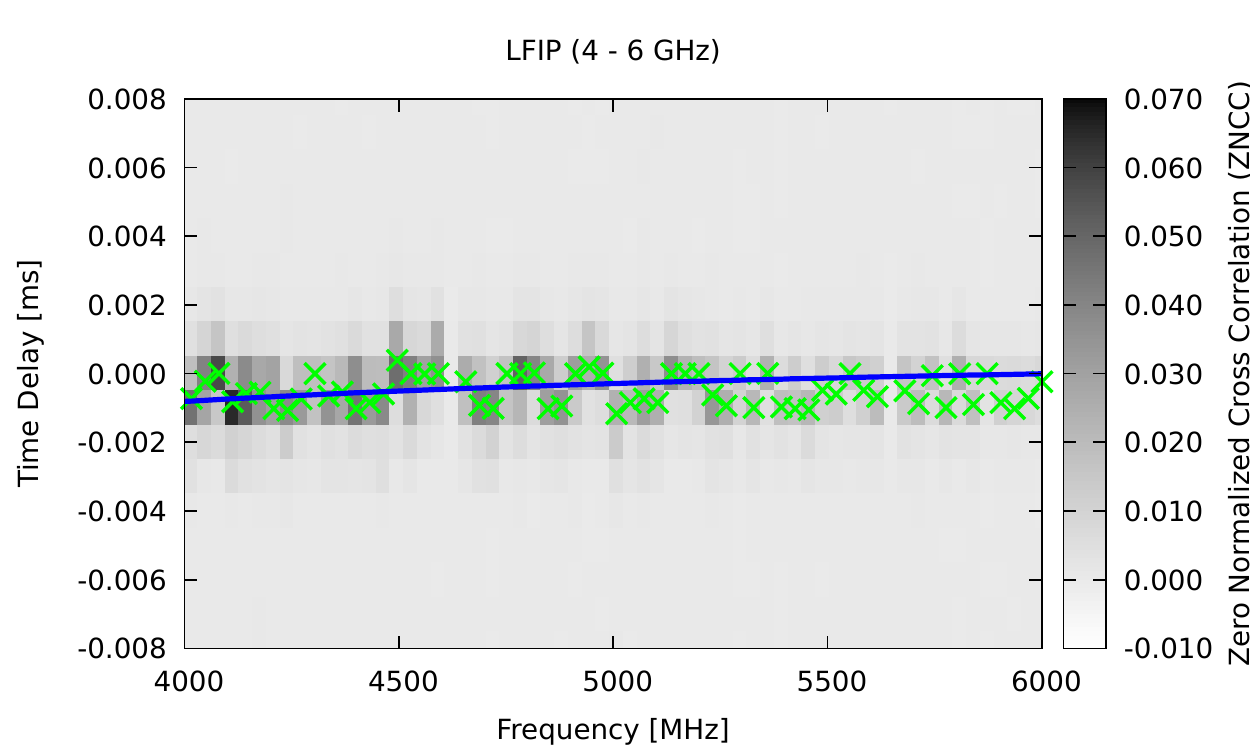}
\caption{Dedispersed Crab pulsar radio data. The green crosses show the peak of the CCF at a frequency $f$ with reference to $f_{1}$ = 4\,GHz. The blue line shows the best dispersion curve. Top plot: data from single pulses at the phase range of the MP dedispersed using DM$_{\rm JB}$ - $\Delta$DM$_{\rm MP}$; middle plot: single pulses from the phase range of the HFIP dedispersed with DM$_{\rm JB}$ - $\Delta$DM$_{\rm HFIP}$; bottom plot: single pulses from the phase range of the LFIP dedispersed using DM$_{\rm JB}$ - $\Delta$DM$_{\rm LFIP}$. All data sets are at 4 - 6 GHz with $\Delta$DM values listed in Table 2.\label{data_applic_II}}
\end{figure}

To determine the correct value for the DM we applied the $\Delta$DM$_{\rm comp}$ values in Table~\ref{method1} separately to the MP, HFIP + LFIP (2 --  4\,GHz) and the MP, HFIP, LFIP (4 -- 6\,GHz). The corresponding results are shown in Figure~\ref{data_applic} and~\ref{data_applic_II}. 
For that purpose the data were dedispersed with the refined DM = DM$_{\rm JB}$ - $\Delta$DM$_{\rm comp}$ and the CCF calculations as described in Section~\ref{least_sqa} were carried out again. The fits displayed in Figure~\ref{data_applic} and~\ref{data_applic_II} (resulting from Stage 2 of the analysis) indicate a resulting delay of nearly zero.
The resulting $\Delta$DM$_{\rm comp}$ values as shown in Table 2 referenced to the respective value of the MP are shown in Table~\ref{delta_dm_components}.
The results of this correction are further discussed in Section~\ref{conc}.

\section{Single Pulses from the HFIP}\label{single_pulses_HFIP}
\begin{figure}
\includegraphics[width=\columnwidth]{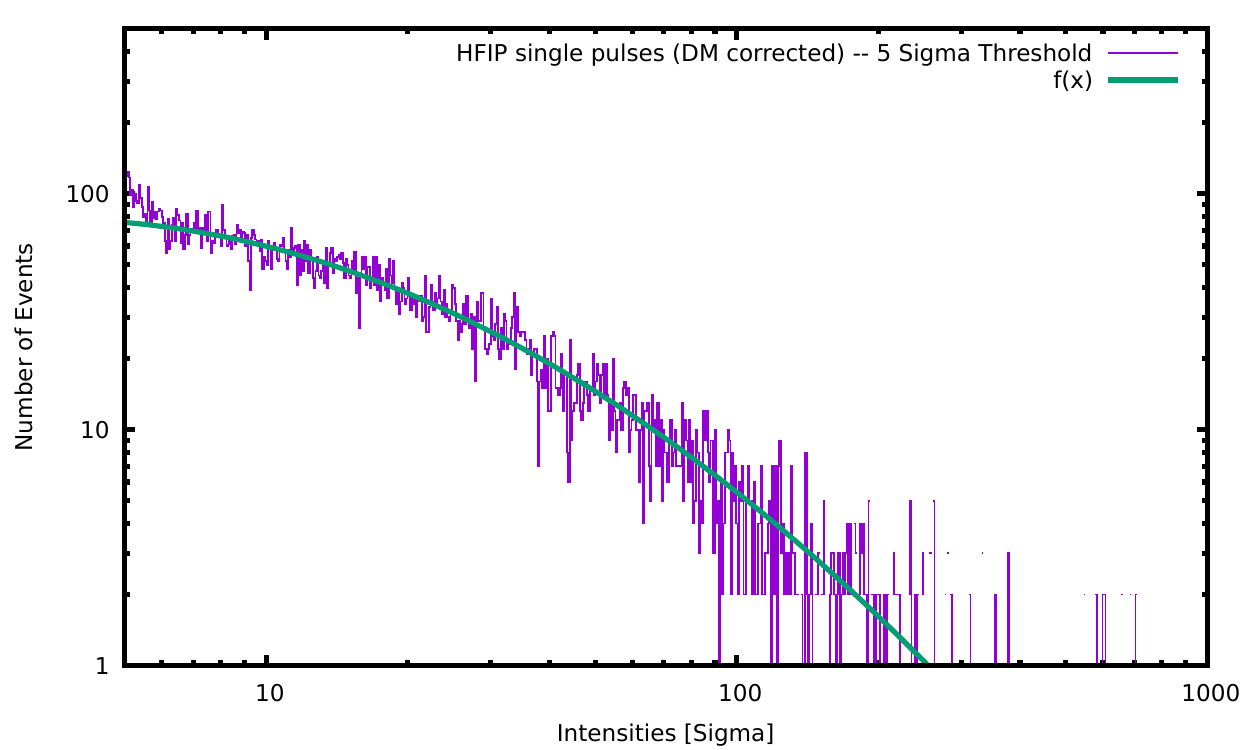}
\caption{Amplitude distribution of HFIP single pulses observed at the full bandwidth of 2\,GHz from 4 to 6 GHz above a treshold of 5\,$\sigma$ fitted with a parabola. \label{amplitude_distributions}}
\end{figure}

\begin{figure}
\includegraphics[width=\columnwidth]{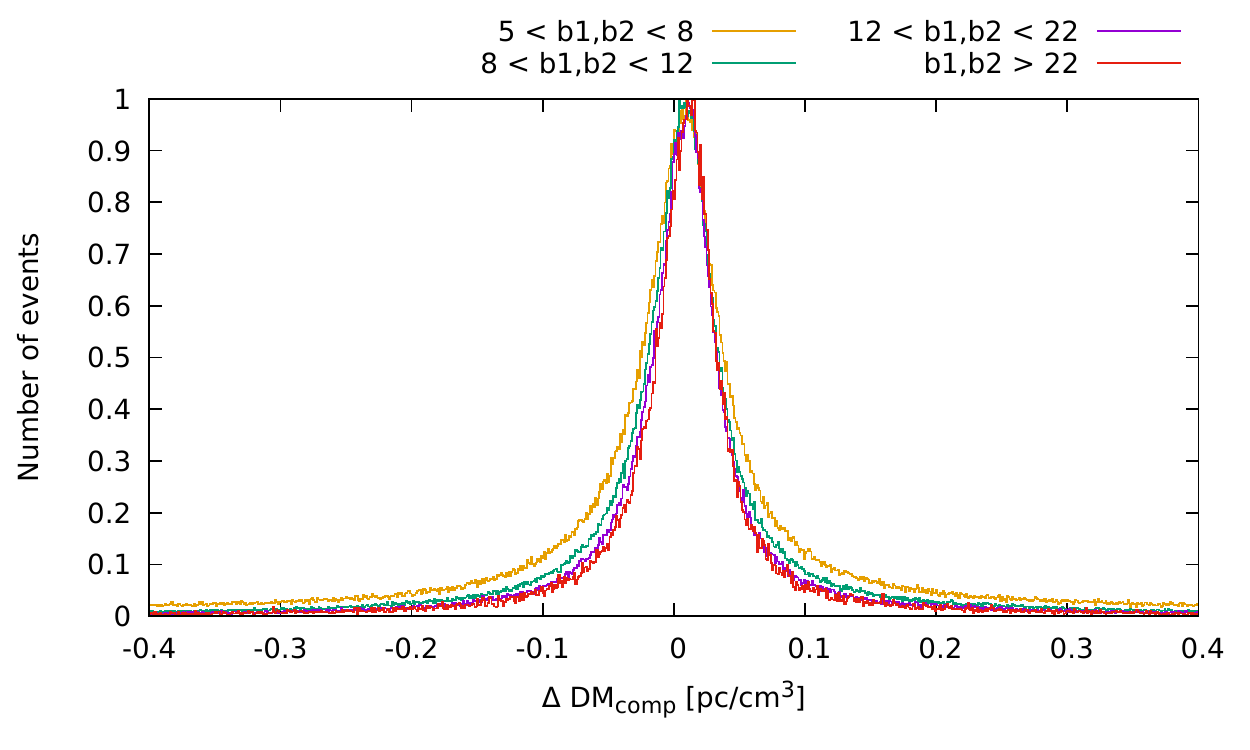}
\caption{The $\Delta$DM distribution of all single pulses larger than 5$\sigma$ in the 4 -- 6 GHz band. Here, b1 and b2 are the peak brightnesses in channel 1 and 2. We see a shift in $\Delta$DM values between very bright and less bright single pulses. \label{HFIP_single_pulses_histos}}
\end{figure}

When performing the calculations described in Section~\ref{freq_time_lag}, we also investigate the influence of the intensity threshold on the resulting distribution of $\Delta$DM$_{\rm comp}$ values.
As shown in Figure~\ref{amplitude_distributions} the intensity distribution of HFIP single pulses can be described by parabolic fit in the log-log space. To examine the influence of the peak brightness of single pulses we split the data by peak brightness into quarters and determine the resulting $\Delta$DM$_{\rm comp}$ values via Method 2 (Section~\ref{freq_time_lag}) in each of the four groups. As described in  Figure~\ref{HFIP_single_pulses_histos} the quantities b1 and b2 are peak brightnesses in the respective channels. For instance, the orange line in Figure~\ref{HFIP_single_pulses_histos} shows the distribution for all single pulses with intensities between 5$\sigma$ and 8$\sigma$.
We applied a logical AND condition for the selections of b1 and b2 shown in Figure~\ref{HFIP_single_pulses_histos}.
In the case of the HFIP we notice a trend with peak brightness, fit each distribution with a Gaussian and find that single pulses with different peak brightness values show different center and width values of their distributions (Table~\ref{fig8_fitting_values}). In other words, brighter single pulses show a noticeably narrower and shifted distribution of $\Delta$DM$_{\rm comp}$ values compared to single pulses with lower peak brightness (Figure~\ref{HFIP_single_pulses_histos}, Table~\ref{fig8_fitting_values}). 

No such difference in distribution of $\Delta$DM values has been found for MP and LFIP single pulses in that frequency band. A more extensive analysis of these results will be provided in an upcoming paper.

\begin{deluxetable*}{c c c c c}
\tablecaption{Gaussian fit parameters for the same cuts in peak brightness as shown in Figure~\ref{HFIP_single_pulses_histos}. \label{fig8_fitting_values}}
\tablecolumns{6}
\tablenum{5}
\tablewidth{0pt}
\tablehead{
\colhead{} &
\colhead{Center$_{\rm Gauss}$} &
\colhead{Error$_{\rm Center}$ } &
\colhead{Width$_{\rm Gauss}$} &
\colhead{Error$_{\rm Width}$}  \\
}
\startdata
5 $<$ b1,b2 $<$ 8 & 0.009 & 0.002 & 0.0253 & 0.002\\
8 $<$ b1,b2 $<$ 12 & 0.009 & 0.001 & 0.0219 & 0.002\\
12 $<$ b1,b2 $<$ 22 & 0.0102 & 0.0005 & 0.0227 & 0.0006\\
b1,b2 $>$ 22 & 0.012 & 0.002 & 0.022 & 0.002\\
\enddata
\end{deluxetable*}

\color{black}
\section{Discussion}\label{disc}
We first focus on the results from the 2 -- 4\,GHz band. 
Method 2 (Section~\ref{freq_time_lag}) produces larger uncertainties for all examined emission components in contrast with Method 1 (Section~\ref{least_sqa}). This is not surprising when one considers that Method 2 contains many more small frequency separations (with larger expected uncertainties) than Method 1.
The upside is that this method can be parallelized more easily and could be computed on the fly without storing the baseband data if sufficient computing power is available. Comparing the results from both methods for each emission component reveals a diverse behavior. 
The results for the MP in the 2 -- 4\,GHz band agree within the range of their corresponding uncertainties. This is not very surprising since the MP is the brightest emission component in this frequency band. We observe a similar result for the HFIP + LFIP emission component which is a bit surprising since we detect a higher number of single bright pulses from the MP in that band (Table~\ref{radio_sum}).
In the 4 -- 6\,GHz band Method 1 also produces results for the MP, HFIP and LFIP emission components with smaller uncertainties whereas Method 2 leads to uncertainties of different extent (mostly standing out for the HFIP emission component). 
The results from both methods agree with each other for each component within the respective error ranges. However, it is noticeable that the $\Delta$DM$_{\rm comp}$ for the HFIP determined with Method 2 is bigger in value than the one determined via Method 1. A similar difference is not seen for the MP and LFIP in that band. One explanation for this might be that the HFIP seems to consist of single bright pulses with different $\Delta$DM$_{\rm comp}$ values (Section~\ref{single_pulses_HFIP}). Hence a determination of one $\Delta$DM$_{\rm comp}$ value for the entire emission component leads to a coarser value than for Method 1. While the latter method seems more suitable for the determination of a $\Delta$DM$_{\rm comp}$ value for all single pulses from one emission component, Method 2 seems more sensitive for the identification of potentially different single pulse populations.

Based on these results we deduce that Method 1 produces statistically more solid results in the case of our data set which is the reason why we apply the $\Delta$DM values derived with this method for the test described in Section~\ref{appl}.

The described methods might be also of interest for studies of FRBs. These radio bursts of extragalactic origin \citep{lorimer_2007,chatterjee_2017} have been observed to show sporadically repeating \citep{spitler_2016,chime_2019} and not repeating radio emission \citep{petroff_2016}. Extensive studies of the repeating FRB known as FRB121102 led to the detection of pulses with time frequency structures \citep{hessels_2019}. According to that study FRB121102 shows subbursts with widths of less than about 1 ms (some of them bearing resemblance in terms of dynamic spectra with radio giant pulses from the Crab pulsar, \citealt{hankins_2007}). In contrast with pulsar studies where a signal to noise maximizing approach is used to determine its DM, \cite{hessels_2019} showed that determining the maximum of the time derivative of peaks of bursts in a frequency averaged profile is a more promising approach to make the individual subbursts visible (a similar approach is also described in \citealt{gajjar_2018}). However, the authors clearly state that they do not determine a DM value for each individual subburst, but a DM for the entire subburst sample. The DM determination approaches presented here can be used as an add on procedure to determine the DM of individual subbursts that are broadband with short widths and show high signal to noise ratios \citep{gourdji_2019}. We expect the CCF to show separate maxima at the DM of the entire burst as well as the DM that aligns subburts in time and optimizes their visibility. We are aware that subbursts from FRBs show a great deal of diversity regarding these quantities, meaning that our approaches cannot be applied on all subbursts from a FRB. However, the correlation based methods could help to determine the DMs of the brightest subbursts and lead to an estimate of potential emission heights in the case of the still unknown source of FRBs. 

\section{Summary of Results and Conclusions}\label{conc}
Here we test for component dependent DM values in the case of the Crab pulsar based on its single pulse emission. The corresponding radio data were recorded at the VLA, covering a frequency range from 2 to 6\,GHz with an instantaneous bandwidth of 2\,GHz. After coherently dedispersing the data we create a time series with 1$\mu$s time resolution and apply a 5$\sigma$ intensity treshold to extract bright single pulses, resulting in the numbers shown in Table~\ref{radio_sum}. We develop two different approaches to search for emission component dependent DM using single pulses: 1) By holding one reference frequency fixed and determining the CCF with all other 63 frequencies from the other channels (Section~\ref{least_sqa}); 2) by determining the CCFs between the frequencies of all 64 channels (Section~\ref{freq_time_lag}). The corresponding results are summarized in Table~\ref{method1}. We apply an  error analysis in the case of both methods (Section~\ref{err_ana}). 
We examine the obtained $\Delta$DM$_{\rm comp}$ values for the MP, HFIP + LFIP (2 -- 4\,GHz), the MP, HFIP and LFIP (4 -- 6\,GHz) emission components and apply the results from Method 1 on the radio data by subtracting them from the initial DM$_{\rm JB}$ = 56.778 pc cm$^{-3}$. Afterwards we dedisperse the radio data again with the corrected values 
(Table~\ref{delta_dm_components}). After that step we apply the CCF algorithm (Section~\ref{least_sqa}) on the newly dedispersed radio data again. The results are shown in Figure~\ref{data_applic} and~\ref{data_applic_II}.

While examining the dependence between the intensity threshold and the $\Delta$DM$_{\rm comp}$ values of single pulses resulting from Method 2, we detect that in the case of HFIP in the 4 -- 6 GHz band single pulses with higher peak brightnesses have different $\Delta$DM$_{\rm comp}$ values than the ones with lower peak brightnesses (Figure~\ref{HFIP_single_pulses_histos}). We fit each of the corresponding distributions with a Gaussian, measuring its center and width (Table~\ref{fig8_fitting_values}). The results of that procedure show that very bright single pulses show higher values of $\Delta$DM$_{\rm comp}$ while the width of the corresponding distribution becomes smaller with increasing peak brightness. We do not see a similar trend in peak brightness for MP and LFIP single pulses in that band.

To examine the influence of profile evolution which is a function of frequency \citep{lentati_2017}, we carry out two additional tests: 1) Look for changes in pulse shape vs.\ frequency since they are not caused by dispersion; 2) look at arrival times vs.\ frequency and examine whether a part does not follow the f$^{-2}$ law as indicated by Equation~\ref{delay}. In both cases we do not find visible results that indicate the influence of profile evolution on our calculations.
However, it is interesting to point out that most of the curves shown in Figures~\ref{data_applic} and ~\ref{data_applic_II} are not flat after correcting with the respective component dependent DM. The exception is the HFIP in the C-band. Comparing the curves from the other two emission components in both bands shows that the curves in Figure~\ref{data_applic} indicate a larger time delay than the ones for the same components shown in Figure~\ref{data_applic_II}. That might indicate that our DM corrections for these components are not entirely correct. However, as a further possibility we cannot exclude the existence of profile evolution in the case of the MP and LFIP either. 
One observational fact that speaks against profile evolution as the reason for the observed difference in DM$_{\rm HFIP}$ is 
that such a detection was already made earlier \citep{hankins_2007,hankins_2015} and at different frequencies than used in the present study. This leads to the conclusion that the observed differences in $\Delta$DM$_{\rm comp}$ of the HFIP emission component are of intrinsic origin. The DM corrections of the MP and LFIP emission components at both bands need further investigation with a potentially larger data set.\color{black}\\

\textbf{Acknowledgments}
NL was supported by NSF award number 1516512. MAM is supported by NSF Physics Frontiers Center award number 2020265 and by NSF AAG award number 2009425.
 The National Radio Astronomy Observatory is a facility of the National Science Foundation operated under cooperative agreement by Associated Universities, Inc..

\begin{deluxetable*}{cccc}
\tablecaption{Position of the peak center $\mu$ and peak width $w$ resulting from fitting the emission components shown in Figure~\ref{pro}.  \label{prof_fits}}
\tablenum{3}
\tablehead{
\colhead{Emission Component} &
\colhead{$\mu$ [Pulse Phase]} &
\colhead{$w$ [Pulse Phase]} &
\colhead{Frequency Band [GHz]}
}
\startdata
 MP          & 0.477 & 0.015 & 2 - 4\\
 HFIP + LFIP & 0.880 & 0.023 & 2 - 4\\
 HFIP        & 0.384 & 0.014 & 3.5 - 4\\
 LFIP        & 0.404 & 0.006 & 3.5 - 4\\
 \hline
 MP          & 0.438 & 0.015 & 4 - 6\\
 HFIP        & 0.808 & 0.065 & 4 - 6\\
 LFIP        & 0.850 & 0.019 & 4 - 6\\
\enddata
\end{deluxetable*}

\begin{deluxetable*}{c c c c c}
\tablecaption{Summary of DM values for emission components applied on the radio data. \label{delta_dm_components}}
\tablecolumns{6}
\tablenum{4}
\tablewidth{0pt}
\tablehead{
\colhead{Emission Component} &
\colhead{DM$_{\rm JB}$ $-$ $\Delta$DM$_{\rm comp}$} &
\colhead{$\Delta$DM$_{\rm comp}$ $-$ $\Delta$DM$_{\rm MP}$} &
\colhead{$\sigma$(DM)} &
\colhead{Frequency Range}  \\
\colhead{} & \colhead{[pc cm$^{-3}$]} &
\colhead{[pc cm$^{-3}$]} & \colhead{[pc cm$^{-3}$]} &
\colhead{GHz}
}
\startdata
MP & 56.7598 & 0 & 0.0002 & 2 - 4\\
HFIP + LFIP & 56.7587 & --0.0011 & 0.0002 & 2 - 4\\
MP & 56.7564 & 0 & 0.0011 & 4 - 6\\
HFIP & 56.7691 & \textbf{0.0127} & 0.0011 & 4 - 6 \\
LFIP & 56.7573 & 0.0009 & 0.0013 & 4 - 6\\
\enddata
\end{deluxetable*}

\bibliographystyle{aasjournal} 
\bibliography{references_cor_resubmission}

\end{document}